# A CFL condition for the finite cell method

Tim Bürchner*[1], Lars Radtke[2] and Philipp Kopp[3]

[1]Chair of Computing in Civil and Building Engineering, Technische Universität München
[2]Chair of Structural Mechanics, Universität Rostock
[3]Chair of Data Science in Civil Engineering, Bauhaus-Universität Weimar

## Abstract

Immersed boundary finite element methods allow the user to bypass the potentially troublesome task of boundary-conforming mesh generation. When combined with explicit time integration, poorly cut elements with little support in the physical domain lead to a severely reduced critical time step size, posing a major challenge for immersed wave propagation simulations. The finite cell method stabilizes cut elements by defining the weak form of the problem also in the fictitious domain, but scaled by a small value $\alpha$. This paper investigates the effect of the finite cell method on the critical time step size for explicit time integration. Starting with an analytical one-degree-of-freedom model, we systematically study the influence of $\alpha$-stabilization on the maximum eigenvalue, and thus on the critical time step size, for corner and sliver cuts. The analysis is complemented by a numerical study of an example with one element and increasing polynomial degree, confirming that the critical time step size does not decrease below a certain limit, even as the cut fraction tends to zero. This lower bound is controlled by the choice of $\alpha$. In higher dimensions, sliver cuts are found to be more detrimental than corner cuts, thus determining the minimum critical time step size. Increasing the polynomial degree has only little effect on this degradation. Based on these observations, we derive an estimate of the minimum critical time step size as a function of $\alpha$, which we use to propose a modified CFL condition for the finite cell method. The validity of this condition is demonstrated on a two-dimensional perforated plate example.

*Keywords:* immersed wave equation, CFL condition, finite cell method, spectral element method, central difference method

## 1. Introduction

The efficient and accurate simulation of acoustic and elastic wave propagation is an essential task in many scientific disciplines, such as seismology [1, 2], medical imaging [3], structural health monitoring [4], or non-destructive testing [5]. While grid-point methods such as the finite difference method (FDM) approximate the underlying partial differential equation (PDE) on a discrete set of grid points, the finite element method (FEM) divides the spatial domain into elements and approximates the solution by a set of basis functions with corresponding coefficients that satisfy the weak form of the PDE [6]. Due to the simplicity of its basic formulation and its efficiency when combined with explicit time integration schemes, FDM and its variants enjoy great popularity in many applications. However, when confronted with complex geometries, the implementation of FDM and its boundary conditions becomes challenging [6]. Standard FEM accurately resolves arbitrary geometries with boundary-conforming meshes and allows to straightforwardly impose all kinds of boundary conditions, e.g., reflecting or absorbing boundaries. Higher-order approaches, such as the spectral element method (SEM) [7, 8], exploit the spectral convergence with respect to the functional interpolation of polynomials to reduce the discretization error. In wave propagation simulation, SEM is widely used due to its favorable properties in terms of accuracy and efficiency [9]. A clever combination of the basis functions' interpolation points and the quadrature rule results in a diagonal mass matrix without deteriorating the solution quality. Integrating in time with explicit methods, such as the central difference method (CDM), then becomes trivial and leads to excellent parallel scaling properties and low memory requirements [10]. The corresponding critical time step size of the time integration can be determined with the CFL condition [11], which takes into account the wave speed and the element size, as well as the Courant number of the CDM, the spectral radius of SEM, and the dimension of the problem [9]. Despite – or rather because of – the underintegration of the mass matrix, SEM shows a $p$ times higher accuracy than classical FEM [12], where $p$ is the polynomial degree of the shape functions, as long as the element distortion is moderate [13]. In addition, the critical time step size increases as a consequence of nodal lumping, since eigenvalues are now approached from below in SEM, whereas they are approached from above in classical FEM. It is important to note that a variety of successful mass lumping approaches exist also for classical finite element formulations.

*tim.buerchner@tum.de, Corresponding author

Generating higher-order boundary-conforming hexahedral meshes remains a tedious task [6]. Immersed boundary methods (IBM) overcome this challenge by embedding the domain of interest $\Omega$ into an extended domain $\Omega^e$ with a simple shape. We use IBM as a general term for immersed boundary finite element methods that is not tied to a specific stabilization method. Moreover, we call $\Omega^f = \Omega^e \setminus \Omega$ the fictitious domain. It is then straightforward to mesh the extended domain with a boundary-conforming mesh (for example, using a Cartesian grid). However, cut elements with little support in the physical domain have adverse effects on the conditioning of the mass and stiffness matrices and the critical time step size during explicit time integration [14, 15]. The finite cell method (FCM) [16, 17] is a stabilization technique for immersed boundary methods that defines the weak form of the problem on both $\Omega$ and $\Omega^f$. To reduce the consistency error, the contribution from $\Omega^f$ is scaled by a stabilization parameter $\alpha$ that is usually chosen close to zero, but large enough to retain definiteness. For specific applications, the finite cell method can often be interpreted as material stabilization. In explicit dynamics, it is equivalent to scaling the density in $\Omega^f$ by $\alpha$, as we will show in Section 2. Advantages of FCM are that its derivation is straightforward, the formulation is independent of the discretization, and its effectiveness has been demonstrated for many linear and nonlinear problems. For Laplace and linear elastic equations, [18] found that the energy norm of the consistency error scales with $\sqrt{\alpha}$. Their analysis also confirms that the finite cell method retains the exponential convergence of $p$-finite element methods until the consistency or quadrature errors dominate. De Prenter et al. [14] give a comprehensive overview of the conditioning and stability of IBMs and discuss possible remedies, focusing on time-independent problems. An isogeometric extension of the finite cell method (IGA-FCM) was introduced by [19] and further studied in [20, 21, 15]. Alternative stabilization techniques include cutFEM [22, 23], aggregated FEM [24, 25], cgFEM [26], or the shifted boundary method [27, 28]. In the context of trimmed IGA, there are extended B-splines [29, 30] and other polynomial extensions [31, 32]. Approaches specifically tailored to the explicit time-integration of immersed hyperbolic problems include eigenvalue stabilization (EVS) [33], local time stepping [34, 35], and implicit-explicit (IMEX) time integration [36–39]. While the references above primarily study the accuracy per degree of freedom, the work in [40] analyzes computation times for SCM and IGA-FCM, which also depend on the sparsity of the system matrices and the time integration scheme.

However, to the authors' knowledge, the influence of the $\alpha$-stabilization on the critical time step size has not been studied in detail. We therefore investigate this relationship by conducting analytical and numerical studies on the FCM-discretized scalar wave equation. For higher polynomial degrees, we consider FCM together with spectral basis functions, a combination that is often referred to as spectral cell method (SCM) [41, 42]. We do not mass-lump cut elements, as it has been shown to significantly degrade the accuracy of SCM simulations [37, 15, 38, 40]. We expect that some results obtained in this study translate to the eigenvalue stabilization approach introduced in [33].

The remainder of the paper is organized as follows. Section 2 introduces the scalar wave equation and its spatial discretization with SCM. Section 3 analytically studies how $\alpha$-stabilization affects the critical time step size on an example with a single linear finite element that is constrained to have only one degree of freedom for any dimension. Section 4 then numerically examines how these results translate to higher polynomial degrees again using one finite element. In Section 5, we propose an estimate of the minimum critical time step size and derive a modified CFL condition for the finite cell method. We verify our results on a two-dimensional perforated plate, where we vary the circle positions to cover as many cut configurations as possible. Based on our findings, we discuss the applicability of the estimate and its limitations. Finally, Section 6 summarizes the paper and outlines the implications for explicit dynamics with FCM in practice.

## 2. Scalar wave equation and finite cell method

Given a density $\rho$, a wave speed $c$, and a time-dependent distributed load $f$, the second-order scalar wave equation for the scalar primary field $\Psi$ is

$$\rho\,\ddot{\Psi} - \nabla \cdot \left(\rho\, c^2\, \nabla\Psi\right) = \rho\, f \quad \text{ in } \Omega \times [0, T], \tag{1}$$

where $\Omega \subset \mathbb{R}^d$ denotes the physical domain with position vector $\boldsymbol{x} \in \mathbb{R}^d$, and $T > 0$ is the final time. We explicitly retain the density in (1) since the material interpretation of the finite cell method for the scalar wave equation diminishes the influence of the fictitious domain by reducing the density there. We also define Neumann and Dirichlet boundary conditions

$$\nabla\Psi \cdot \boldsymbol{n} = \bar{v}_n \quad \text{ on } \Gamma^{\mathrm{N}}, \quad \text{ and } \quad \Psi = \bar{\Psi} \quad \text{ on } \Gamma^{\mathrm{D}}, \tag{2}$$

where $\boldsymbol{n}$ is the outward-pointing unit normal vector and $\Gamma^{\mathrm{N}} \cup \Gamma^{\mathrm{D}} = \partial\Omega$ forms the domain boundary, such that $\Gamma^{\mathrm{N}} \cap \Gamma^{\mathrm{D}} = \emptyset$. We finalize the problem definition by introducing the initial states $\Psi(\boldsymbol{x}, 0) = \Psi_0(\boldsymbol{x})$ and $\dot{\Psi}(\boldsymbol{x}, 0) = \dot{\Psi}_0(\boldsymbol{x})$.

Following the philosophy of immersed boundary finite element methods, we embed the physical domain $\Omega$ into an easy-to-mesh extended domain $\Omega^{\mathrm{e}}$ (usually a Cartesian grid). The finite cell method introduces the indicator function

$$\alpha^{\mathrm{FCM}}(\boldsymbol{x}) = \begin{cases} 1 & \text{for } \boldsymbol{x} \in \Omega, \\ \alpha & \text{else}, \end{cases} \tag{3}$$

to distinguish between $\Omega$ and $\Omega^{\mathrm{f}}$, where $\alpha \in [0, 1]$ is a stabilization parameter. Multiplying (1) by a test function $\delta\Psi$, integrating by parts, multiplying by $\alpha^{\mathrm{FCM}}$, and substituting the Neumann boundary condition results in the stabilized weak form of the scalar wave equation:

$$\int_{\Omega^{\mathrm{e}}} \alpha^{\mathrm{FCM}}\, \rho\, \ddot{\Psi}\, \delta\Psi + \alpha^{\mathrm{FCM}}\, \rho\, c^2\, \nabla\Psi \cdot \nabla\delta\Psi \, \mathrm{d}\Omega^{\mathrm{e}} = \int_{\Omega^{\mathrm{e}}} \alpha^{\mathrm{FCM}}\, \rho\, f\, \delta\Psi \, \mathrm{d}\Omega^{\mathrm{e}} + \int_{\Gamma^{\mathrm{N}}} \alpha^{\mathrm{FCM}}\, \rho\, c^2\, \bar{v}_n\, \delta\Psi \, \mathrm{d}\Gamma^{\mathrm{N}}. \tag{4}$$

Setting $\alpha = 0$ recovers the original solution in $\Omega$, but leads to a loss of definiteness since there is no equation to determine $\Psi$ inside $\Omega^{\mathrm{f}}$. Selecting a small but nonzero value for $\alpha$ recovers a unique solution in $\Omega^{\mathrm{f}}$, at the cost of limiting the accuracy by introducing a consistency error [18]. Moreover, if $\Gamma^D \not\subset \partial\Omega^e$, strong boundary conditions cannot be imposed the usual way, and one must instead use integral methods. While the weak imposition of Dirichlet boundary conditions is not the main focus of this paper, we briefly discuss the consequences at the end of Section 4.

To construct finite element spaces for discretizing (4), we employ Lagrange polynomials with Gauss-Lobatto-Legendre (GLL) interpolation points as shape functions. Let $\xi_{0,k}^{\mathrm{Lo},p-1}$, $k \in \{1, \ldots, p-1\}$ denote the roots of the Lobatto polynomials of order $p-1$. Then, the $p+1$ GLL points on a reference interval $[-1, 1]$ are

$$\xi_j = \begin{cases} -1 & \text{if } j = 1 \\ \xi_{0,j-1}^{\mathrm{Lo},p-1} & \text{if } 2 \leq j < p+1 \\ 1 & \text{if } j = p+1 \end{cases}, \tag{5}$$

with the corresponding univariate Lagrange polynomials defined as

$$N_i^{\mathrm{Lag},p}(\xi) = \prod_{j=1, j\neq i}^{p+1} \frac{\xi - \xi_j}{\xi_i - \xi_j} \quad \text{ for } i \in \{1, \ldots, p+1\}. \tag{6}$$

We then construct $(p+1)^d$ multivariate shape functions on the $d$-dimensional reference element $\boldsymbol{\xi} \in [-1,1]^d$ via the tensor product of (6):

$$N^p_{i_1,\dots,i_d}(\boldsymbol{\xi}) = \prod_{k=1}^{d} N^{\text{Lag},p}_{i_k}(\xi_k) \quad \text{for } i_k \in \{1,\dots,p+1\}. \tag{7}$$

By mapping the shape function defined in (7) for each element from the reference to the global coordinate system and connecting them across element boundaries, we obtain the basis functions $N_i(\boldsymbol{x})$, with $i \in \{1,\dots,n^{\text{dof}}\}$, where $n^{\text{dof}}$ is the total number of degrees of freedom. Applying the Bubnov-Galerkin method together with the method of lines, we use $N_i(\boldsymbol{x})$ to construct the discrete test functions $\delta\Psi^h$ and the time-dependent solution interpolation $\Psi^h$:

$$\delta\Psi(\boldsymbol{x}) \approx \delta\Psi^h(\boldsymbol{x}) = \sum_{i=1}^{n^{\text{dof}}} N_i(\boldsymbol{x})\,\delta\Psi_i = \mathbf{N}(\boldsymbol{x})\delta\boldsymbol{\Psi}, \tag{8}$$

$$\Psi(\boldsymbol{x},t) \approx \Psi^h(\boldsymbol{x},t) = \sum_{i=1}^{n^{\text{dof}}} N_i(\boldsymbol{x})\,\Psi_i(t) = \mathbf{N}(\boldsymbol{x})\boldsymbol{\Psi}(t), \tag{9}$$

where $\mathbf{N}(\boldsymbol{x})$ is the row vector of basis functions, and $\boldsymbol{\Psi}$ is the column vector of time-dependent solution coefficients. Substituting (8) and (9) into the weak form (4), yields the semi-discrete wave equation

$$\mathbf{M}\,\ddot{\boldsymbol{\Psi}} + \mathbf{K}\,\boldsymbol{\Psi} = \mathbf{F}, \tag{10}$$

where $\mathbf{M}$, $\mathbf{K}$, and $\mathbf{F}$ denote the global mass matrix, stiffness matrix, and load vector. An essential component of immersed finite elements is the numerical quadrature of the discontinuous integrands. While Sections 3 and 4 feature simple cut geometries permitting an exact evaluation of the finite element integrals, Section 5 describes the adaptive numerical quadrature of (4) using a combination of GLL and Gauss-Legendre points.

The time integration of (10) is commonly performed by the explicit CDM due to its simplicity. As detailed in the introduction, the critical time step size and, hence, the largest eigenvalue are a major obstacle for combining most immersed boundary methods with explicit time integration schemes. This is because cut elements with little support in the physical domain lead to large eigenvalues, necessitating the use of stabilization methods. Let $\lambda_{\max}(\mathbf{K},\mathbf{M})$ denote the largest eigenvalue of the generalized eigenvalue problem associated with the stiffness and mass matrices. Then, the critical time step size of the CDM [43] is given by

$$\Delta t_{\text{crit}} = \frac{2}{\sqrt{\lambda_{\max}(\mathbf{K},\mathbf{M})}}. \tag{11}$$

To summarize, selecting a small positive value for $\alpha$ recovers definiteness of the extended system and imposes upper limits to the condition numbers of the mass and stiffness matrices, and to the largest eigenvalue of the corresponding generalized eigenvalue problem [14]. This $\alpha$-stabilization introduces a consistency error that scales with $\sqrt{\alpha}$ in the energy norm for Laplace and Helmholtz problems [18]. We can interpret $\alpha$-stabilization from a continuum mechanics point of view as scaling the density of the material by $\alpha$ in the fictitious domain while leaving the wave speed untouched [17].

## 3. Analytic investigations

### Problem statement

We start with a single-element setup on $\Omega^e = [0,1]^d$, where we consider two different configurations: corner cuts and sliver cuts. The function space is reduced to a single degree of freedom for arbitrary dimensions by using linear shape functions combined with Dirichlet boundary conditions on the sides sharing a node with the origin. This leaves unconstrained only the shape function associated to the node on the corner opposite to the origin:

$$N(\boldsymbol{x}) = \prod_i x_i.$$

We then introduce the cut position $\chi$, such that $\Omega = [0, \phi]^d$ for corner cuts and $\Omega = [0, \phi] \times [0, 1]^{d-1}$ for sliver cuts, where in both cases $\Omega^f = [0, 1]^d \setminus \Omega$. Figure 1 illustrates this setup for $d = 1$ to $d = 3$ and both cut configurations. In both cases, the volume fraction $\phi(\chi) = \frac{V(\Omega)}{V(\Omega^e)}$ approaches zero when $\chi \to 0$, where $V(\Omega)$ and $V(\Omega^e)$ denote the measure of $\Omega$ and $\Omega^e$.

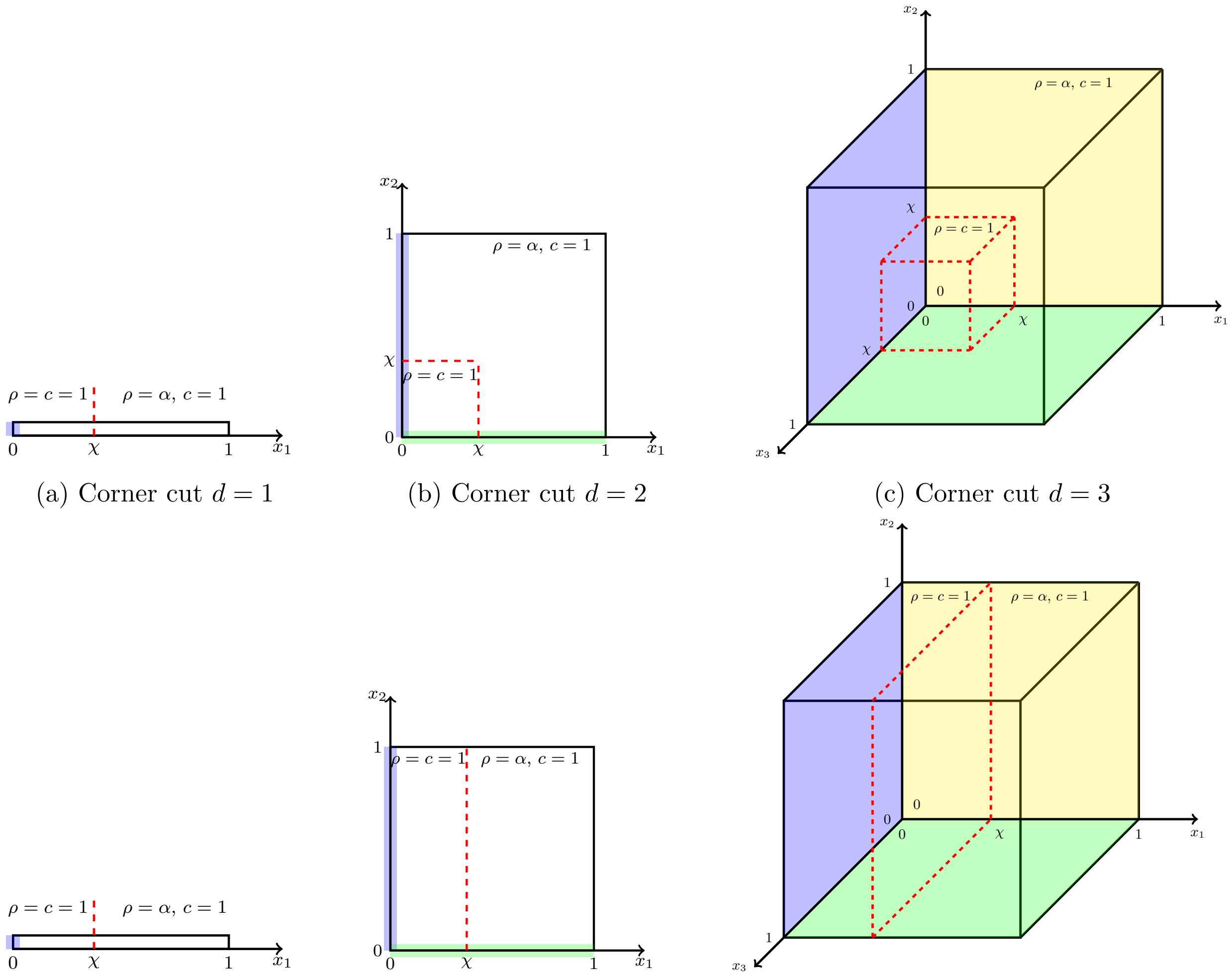

(a) Corner cut $d = 1$ (b) Corner cut $d = 2$ (c) Corner cut $d = 3$

(d) Sliver cut $d = 1$ (e) Sliver cut $d = 2$ (f) Sliver cut $d = 3$

Figure 1: Single degree of freedom systems for corner cuts (top) and sliver cuts (bottom). Dirichlet boundary conditions for faces with $x_1 = 0$, $x_2 = 0$, and $x_3 = 0$ are marked in transparent blue, green, and yellow.

The reduction to a single unknown reduces the mass and stiffness matrices to scalar values, allowing us to solve the problem analytically. The result generates insight into $\alpha$-stabilization and the way it imposes a lower limit on the critical time step size, depending on the type and severity of the cut. To this end, we define $\rho = c = 1$ in the physical domain, and in the fictitious domain we adopt the interpretation of FCM as density scaling, i.e., we set $\rho = \alpha$ and leave the wave speed unaffected ($c = 1$). Following the discretization procedure outlined in the previous section, we obtain the following mass and stiffness expressions

$$M = \int_\Omega N^2(\boldsymbol{x})\, d\Omega + \int_{\Omega^\mathrm{f}} \alpha\, N^2(\boldsymbol{x})\, \mathrm{d}\Omega^\mathrm{f} = \int_\Omega (1-\alpha)\, N^2(\boldsymbol{x})\, d\Omega + \int_{\Omega^\mathrm{e}} \alpha\, N^2(\boldsymbol{x})\, \mathrm{d}\Omega^\mathrm{e}, \tag{12}$$

$$\begin{aligned} K &= \int_\Omega \nabla N(\boldsymbol{x}) \cdot \nabla N(\boldsymbol{x})\, d\Omega + \int_{\Omega^\mathrm{f}} \alpha\, \nabla N(\boldsymbol{x}) \cdot \nabla N(\boldsymbol{x})\, \mathrm{d}\Omega^\mathrm{f} \\ &= \int_\Omega (1-\alpha)\, \nabla N(\boldsymbol{x}) \cdot \nabla N(\boldsymbol{x})\, d\Omega + \int_{\Omega^\mathrm{e}} \alpha\, \nabla N(\boldsymbol{x}) \cdot \nabla N(\boldsymbol{x})\, \mathrm{d}\Omega^\mathrm{e}, \end{aligned} \tag{13}$$

with the corresponding eigenvalue $\lambda = \lambda_\mathrm{max} = \frac{K}{M}$.

## Results

In 1D, corner cuts and sliver cuts are equivalent. The basis function is $N(x) = x$ and the gradient is $\nabla N(x) = 1$. We can express the mass and stiffness as a function of $\alpha$ and $\chi$:

$$M(\chi, \alpha) = \int_0^\chi (1-\alpha)\, x^2\, \mathrm{d}x + \int_0^1 \alpha\, x^2\, \mathrm{d}x = \frac{1}{3}((1-\alpha)\, \chi^3 + \alpha), \tag{14}$$

$$K(\chi, \alpha) = \int_0^\chi (1-\alpha)\, \mathrm{d}x + \int_0^1 \alpha\, \mathrm{d}x = (1-\alpha)\, \chi + \alpha. \tag{15}$$

The system has one eigenvalue, which is

$$\lambda(\chi, \alpha) = \frac{K}{M} = 3\, \frac{(1-\alpha)\, \chi + \alpha}{(1-\alpha)\, \chi^3 + \alpha}. \tag{16}$$

Figure 2 shows $M$, $K$, and $\lambda$ depending on $\chi$ and $\alpha$, when the cut parameter and stabilization are varied between $\chi \in [10^{-16}, 10^0]$ and $\alpha \in [10^{-16}, 10^0]$ on a logarithmically spaced grid of $801 \times 801$ points.

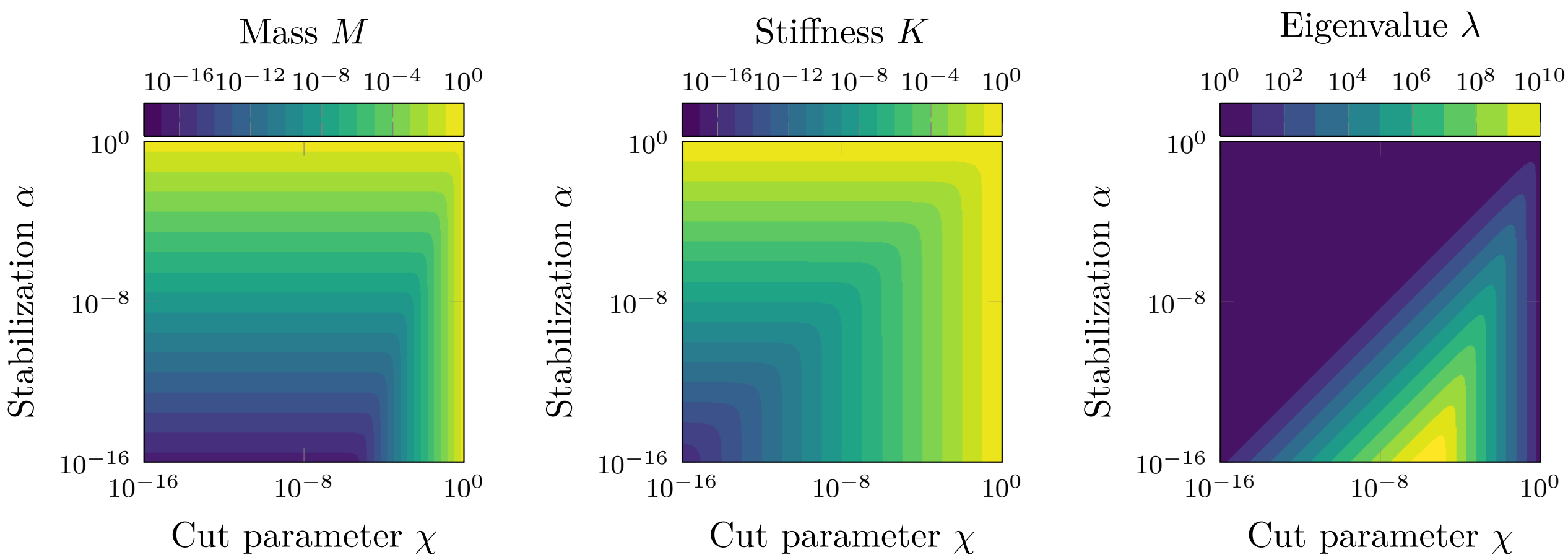

Figure 2: Scalar mass $M$ (left), stiffness $K$ (middle), and eigenvalue $\lambda$ (right) of the one-dimensional system with one degree of freedom.

Table 1: Generalization of the eigenvalue expression

| | Corner cut | Sliver cut |
|---|---|---|
| 1D | $M = \frac{1}{3}((1-\alpha)\,\chi^3 + \alpha)$<br>$K = (1-\alpha)\,\chi + \alpha$<br>$\lambda = 3\,\frac{(1-\alpha)\,\chi+\alpha}{(1-\alpha)\,\chi^3+\alpha}$ | |
| $d$D | $M = \frac{1}{3^d}\,((1-\alpha)\,\chi^{3d} + \alpha)$<br>$K = \frac{d}{3^{d-1}}\,((1-\alpha)\,\chi^{3d-2} + \alpha)$<br>$\lambda = 3\,d\,\frac{(1-\alpha)\,\chi^{3d-2}+\alpha}{(1-\alpha)\,\chi^{3d}+\alpha}$ | $M = \frac{1}{3^d}((1-\alpha)\,\chi^3 + \alpha)$<br>$K = \frac{1}{3^{d-1}}((1-\alpha)\,((d-1)\,\chi^3+\chi) + d\,\alpha)$<br>$\lambda = 3\,\frac{(1-\alpha)\,((d-1)\,\chi^3+\chi)+d\,\alpha}{(1-\alpha)\chi^3+\alpha}$ |

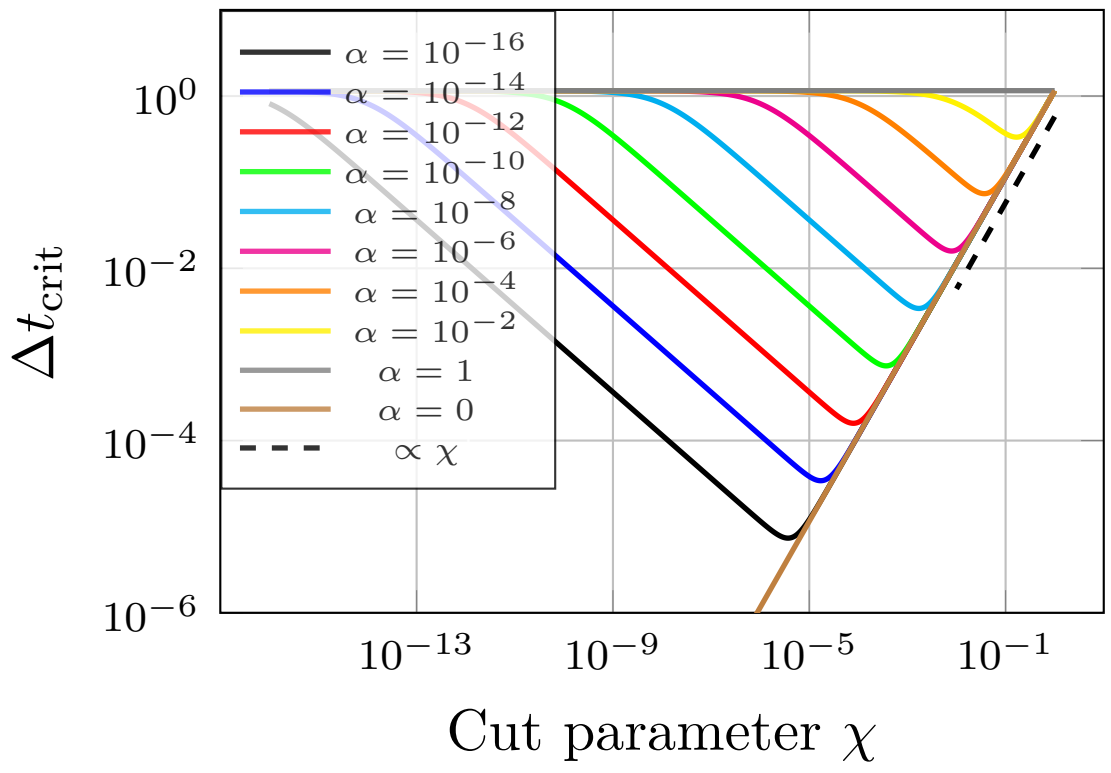

(a) $d = 1$

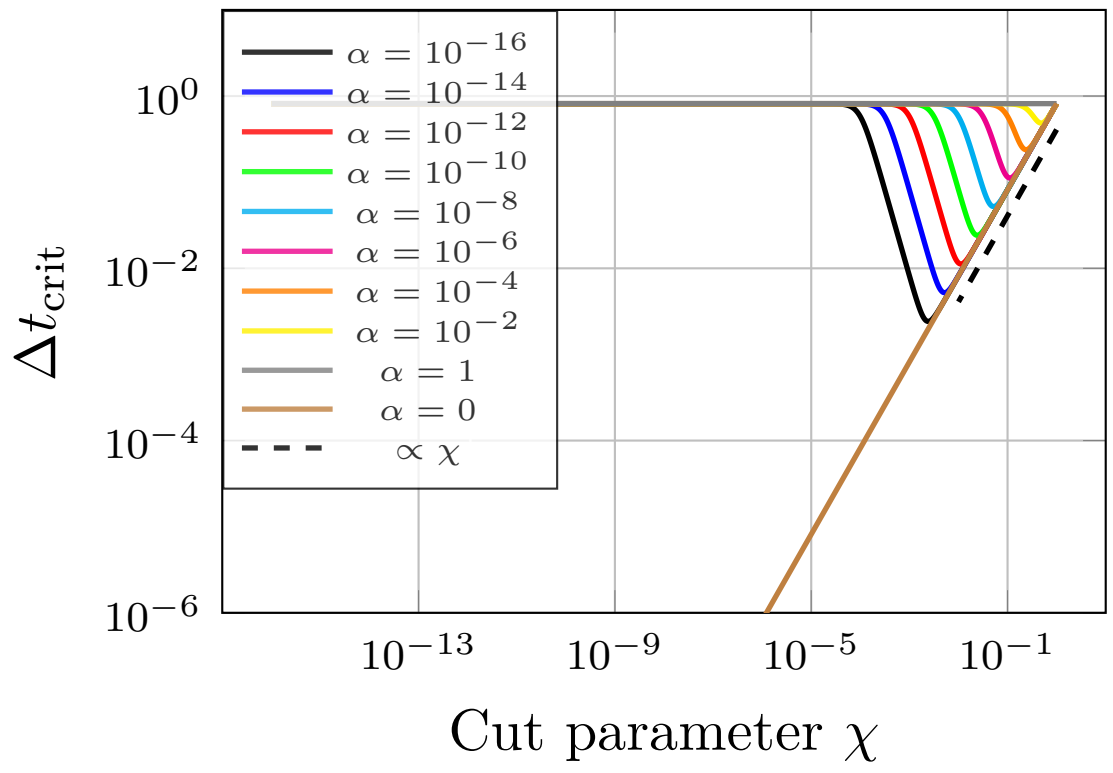

(b) $d = 2$

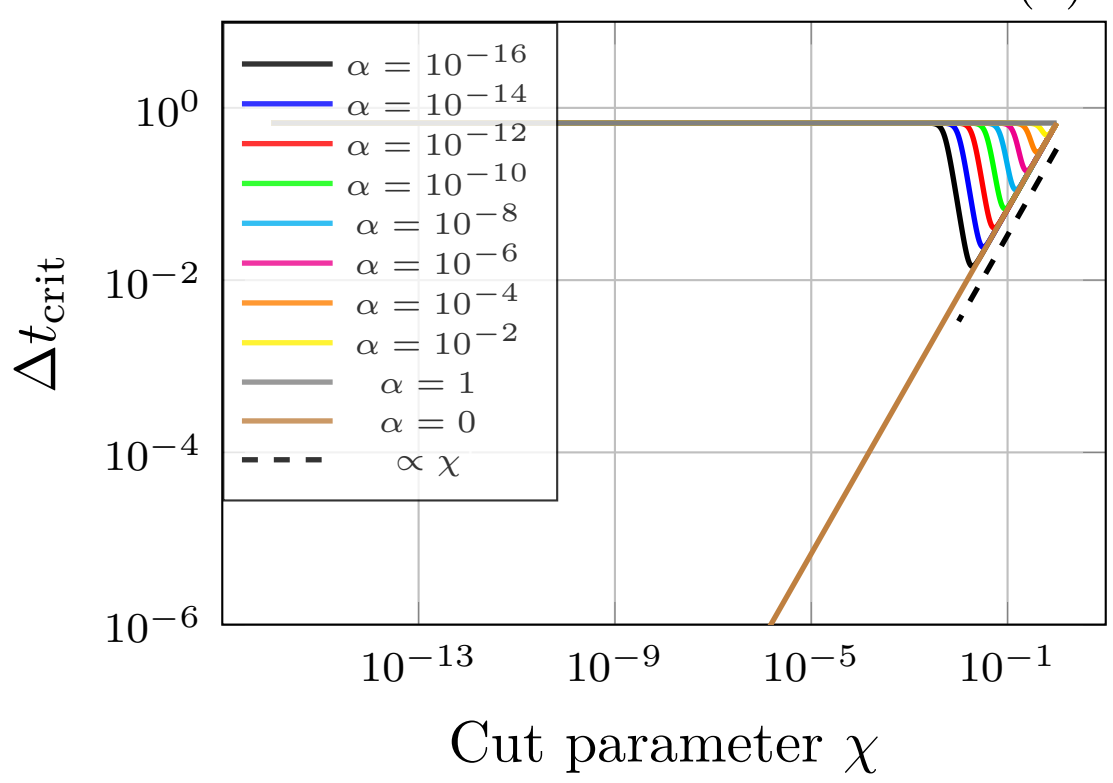

(c) $d = 3$

Figure 3: Corner cuts: Critical time step size $\Delta t_{\text{crit}}$ of system with one degree of freedom for $d = 1, 2$, and 3.

The expressions for $M$, $K$, and $\lambda$ can be extended to any dimension following the same procedure as in one dimension. Table 1 comprises these analytic expressions with respect to $\chi$ and $\alpha$ for any dimension $d$. Figure 3 and Figure 4 show the critical time step sizes with respect to the cut parameter $\chi$ for nine different values of $\alpha$ for corner and sliver cuts in one, two, and three dimensions. Finally, Figure 5 shows the lower limit of the critical time step size, denoted as $\Delta t_{\text{crit,min}}(\alpha)$, for the two cut configurations. The minimum critical time step size for a given $\alpha$ is defined as the minimum value

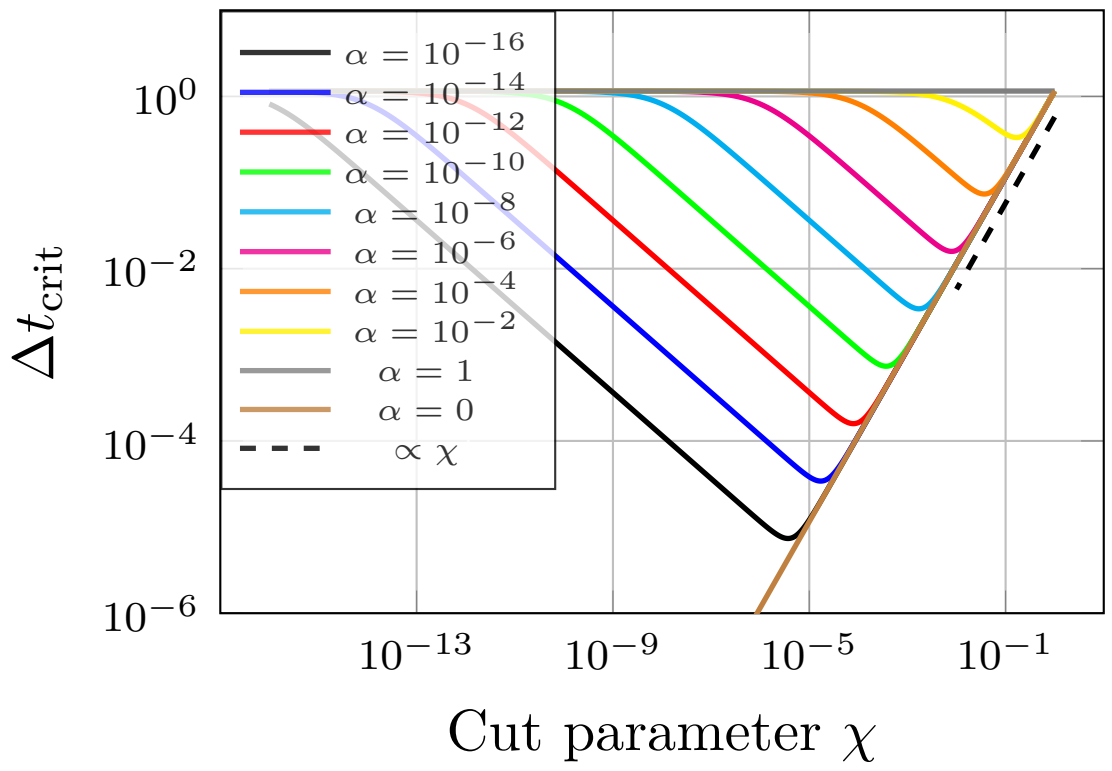

(a) $d = 1$

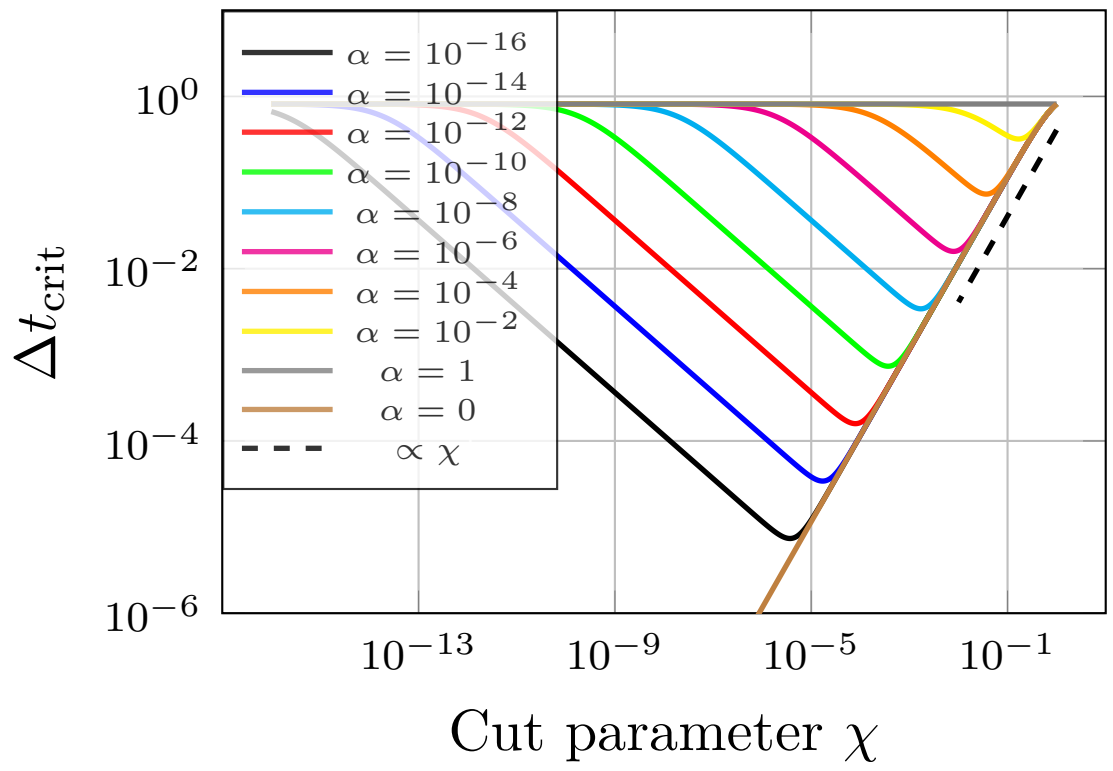

(b) $d = 2$

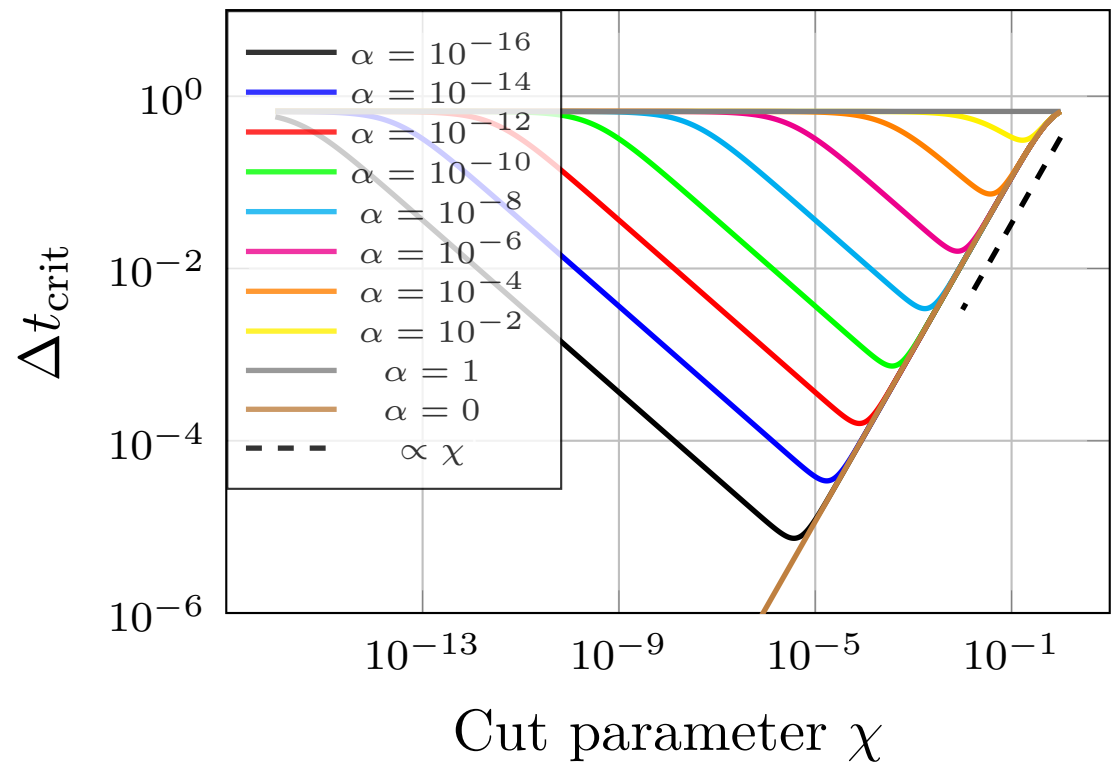

(c) $d = 3$

Figure 4: Sliver cuts: Critical time step size $\Delta t_{\text{crit}}$ of system with one degree of freedom for $d = 1, 2,$ and 3.

of $\Delta t_{\text{crit}}(\chi, \alpha)$ when varying $\chi$:

$$\Delta t_{\text{crit,min}}(\alpha) = \min_{\chi} \Delta t_{\text{crit}}(\chi, \alpha). \tag{17}$$

In the following, $\Delta t_{\text{crit,min}}(\alpha)$ is approximated by the smallest value of $\Delta t_{\text{crit}}(\chi, \alpha)$ of the sampling with respect to $\chi$.

## Discussion

We begin our analysis by considering several edge cases. For the unstabilized system ($\alpha = 0$), the eigenvalues corresponding to the two types of cut configurations are:

$$\lambda^{\text{cc}}(\chi, \alpha = 0) = 3d\chi^{-2} \qquad \text{(corner cut)}, \tag{18}$$

$$\lambda^{\text{sc}}(\chi, \alpha = 0) = 3(d-1) + 3\chi^{-2} \qquad \text{(sliver cut)}. \tag{19}$$

In both cases, the eigenvalue scales quadratically with the inverse of $\chi$, independently of the spatial dimension $d$. As a result, the critical time step size reduces linearly with $\chi$. This recovers the well-known behavior for traditional finite elements when reducing the element size [21]. For $\alpha = 0$ and $\chi = 0$, the eigenvalue is undefined, since both the mass and the stiffness are zero. The other edge cases are:

- $\alpha = 1$, $\chi$ arbitrary: no density reduction in $\Omega^f$
- $\chi = 1$, $\alpha$ arbitrary: uncut/full element

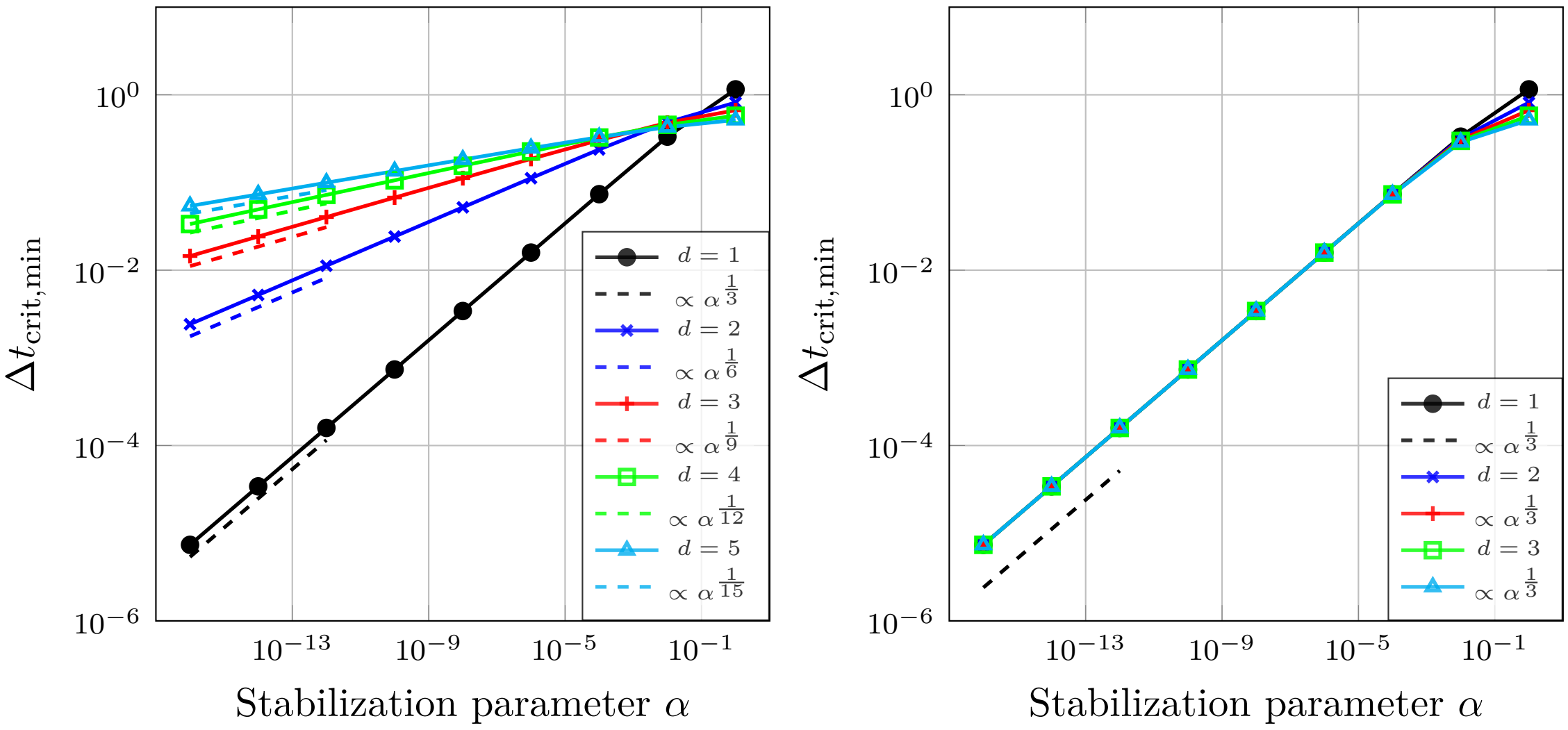

Figure 5: Minimum critical time step sizes $\Delta t_{\text{crit,min}}$ with respect to varying $\alpha$ for dimensions $d = 1$ to $d = 5$ for corner cuts (left) and sliver cuts (right).

- $\chi = 0$, $\alpha \neq 0$ stabilized fictitious/empty element.

For these, we obtain a constant eigenvalue

$$\lambda(\alpha = 1, \chi) = \lambda(\alpha \neq 0, \chi = 0) = \lambda(\alpha, \chi = 1) = \lambda^{\text{full}} = 3d, \tag{20}$$

which we denote as the eigenvalue of a full element $\lambda^{\text{full}}$. This value is independent of both the cut configuration and the stabilization parameter.

In the general case with $\alpha \neq 0$, the material stabilization imposes an upper limit on the eigenvalue, and, correspondingly, enforces a lower limit on the critical time step size. Figures 3 and 4 show that the critical time step size initially decreases when reducing $\chi$, but eventually reaches a minimum and then increases again. The cut configuration, the stabilization parameter $\alpha$, and the dimension $d$ determine the location and value of the minimum critical time step size $t_{\text{crit,min}}$. For corner cuts, this minimum becomes less severe in higher dimensions. As the left plot of Figure 5 shows, the minimum critical time step size scales with $\alpha^{1/(3d)}$; a result we confirm analytically via an asymptotic analysis for small $\alpha$ in Appendix A. In contrast, for sliver cuts the minimum critical time step size is independent of the dimension. As demonstrated in the right plot of Figure 5, it scales with $\alpha^{1/3}$, which we validate again through an asymptotic analysis in Appendix B. Lastly, we point out that only for sliver cuts along one axis the critical time step size is independent of the dimension. For cuts featuring multiple axes (e.g., an edge cut in 3D), the severity of the critical time step size reduces with higher dimensions.

## 4. Numerical Investigations

### Problem statement

The initial analytical study using a single degree of freedom example reveals that the issue of small critical time step sizes is more severe for sliver cuts in higher dimensions. Building on this observation, we investigate the behavior of this cut configuration for multi-degree-of-freedom systems employing high-order shape functions. This extension is motivated by the fact that the solution of wave propagation problems benefits from the spectral convergence properties of higher-order polynomial approximations of smooth functions. In practice, polynomial degrees $p = 4$ or $p = 5$ are commonly used in the spectral element method [9]. The following numerical study examines how increasing the polynomial degree affects the critical time step size. We systematically vary $\chi$, $\alpha$, and $p$ for dimensions $d = 1$ to $d = 3$, with a particular focus on sliver-type geometries. We use the setup of the single degree of freedom example introduced in the previous section (Figure 1), but we impose homogeneous Neumann boundary conditions on all sides, resulting in $(p+1)^d$ unconstrained basis functions. The mass and stiffness matrices now have $(p+1)^d \times (p+1)^d$ entries defined as

$$M_{ij} = \int_{\Omega} (1-\alpha)\, N_i(\boldsymbol{x})\, N_j(\boldsymbol{x})\, d\Omega + \int_{\Omega^{\mathrm{e}}} \alpha\, N_i(\boldsymbol{x})\, N_j(\boldsymbol{x})\, \mathrm{d}\Omega^{\mathrm{e}}, \tag{21}$$

$$K_{ij} = \int_{\Omega} (1-\alpha)\, \nabla N_i(\boldsymbol{x}) \cdot \nabla N_j(\boldsymbol{x})\, d\Omega + \int_{\Omega^{\mathrm{e}}} \alpha\, \nabla N_i(\boldsymbol{x}) \cdot \nabla N_j(\boldsymbol{x})\, \mathrm{d}\Omega^{\mathrm{e}}. \tag{22}$$

Each of the integrals in (21) and (22) are evaluated using Gauss-Legendre quadrature with $p+1$ integration points per spatial direction. Consequently, even for a full (uncut) element, i.e. $\chi = 1$, the mass matrix remains fully populated, as we employ exact integration rather than nodal lumping. The critical time step is now governed by $\lambda_{\max}(\mathbf{K}, \mathbf{M})$, i.e. the largest generalized eigenvalue, which we compute using the SciPy function `scipy.sparse.linalg.eigsh` [44].

### Results

Figure 6a shows the critical time step size for $\alpha = 10^{-4}$ (black) and $\alpha = 10^{-8}$ (blue), both evaluated for polynomial degrees $p = 1$ to $p = 5$. The cut parameter is sampled over the interval $\chi \in [10^{-8}, 10^0]$ with 161 logarithmically spaced values. Like in the previous section, we extract an approximation to the minimum critical time step size from these samples. Figure 6b shows the ratio between the minimum and the full (uncut) critical time step when reducing $\alpha$. Here, we use as full critical time step the exactly integrated, consistent mass matrix:

$$\Delta t_{\text{crit,full}} = \Delta t_{\text{crit,full-c}} = \Delta t_{\text{crit}}(\chi = 1). \tag{23}$$

The results shown are based on the one-dimensional setup. For the sliver cut, they are almost identical in higher dimensions, differing only by a factor close to one.

As an alternative to the uncut consistent mass matrix as a reference, one could choose the critical time step size of the diagonal mass matrix obtained by nodal lumping, which we denote as $\Delta t_{\text{crit,full-l}}$. Both choices do not depend on $\alpha$, and for all configurations investigated in this section, the lumped critical time step size is at most three times larger than that of the consistent mass matrix. Moreover, this factor decreases for higher polynomial degrees.

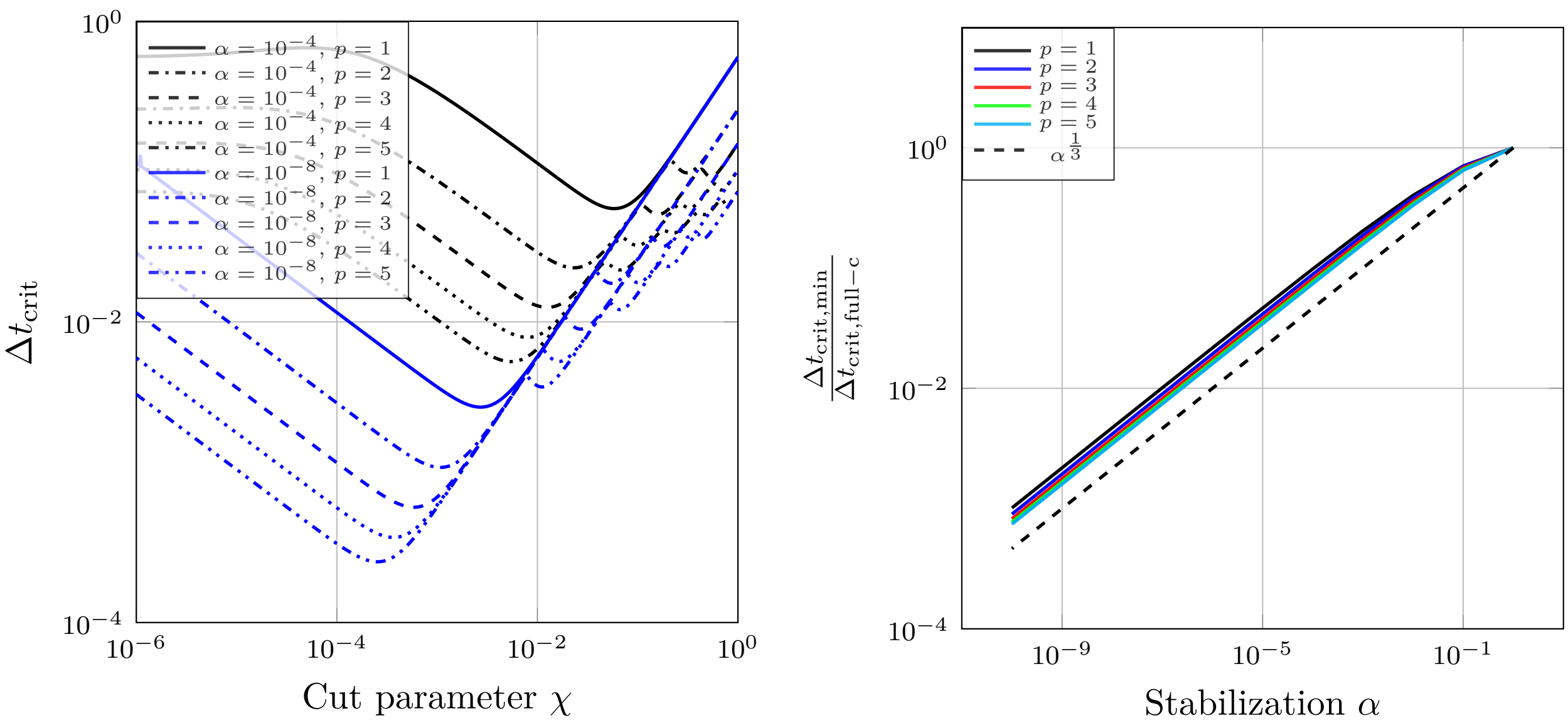

(a) Dependency on the cut position

(b) Ratio between stabilized cut and uncut cells

Figure 6: Critical time step size in one dimension with high-order discretizations. The results are almost identical for sliver cuts in higher dimensions.

### Discussion

Compared to the single-degree-of-freedom example in the previous section, the use of higher order polynomials introduces additional local minima to the time step size as a function of the cut position. However, the overall trend remains the same: the critical time step size initially decreases with the cut position until a global minimum is reached and then increases until $\chi = 0$. The critical time step size remains unbounded in the absence of stabilization. Increasing $\alpha$ has two effects: it raises the minimum critical time step size and shifts the location of the global minimum towards larger values of the cut parameter, independent of the polynomial degree. In addition, the ratio between minimum and full critical time step shown in Figure 6b scales only weakly with the polynomial degree.

We also studied the weak enforcement of Dirichlet boundary conditions using Nitsche's or a penalty method. The penalty parameter was chosen inversely proportional to the cut ratio [21]. In this case, $\alpha$-stabilization does not impose a lower bound on the critical time step anymore. Similarly, the use of Hinton–Rock–Zienkiewicz (HRZ) lumping [45] for diagonalizing the element mass matrices of cut elements also eliminates the effect of $\alpha$-stabilization. As the cut position approaches zero, the critical time step size reduces to zero in both cases.

## 5. A modified FCM-CFL condition

### Minimum critical time step size estimate

Based on the previous findings, we propose a modified Courant–Friedrichs–Lewy (CFL) condition [11] tailored to the finite cell method. This condition provides a generally applicable estimate for the range of stable time steps. As Figure 6b shows, the minimum critical time step size scales with $\alpha^{\frac{1}{3}}$, which is consistent with the analytical result obtained for the single-degree-of-freedom model in Section 3. Accordingly, we propose the following lower-bound estimate:

$$\Delta t_{\mathrm{crit,cfl,fcm}} = \alpha^{\frac{1}{3}}\, \Delta t_{\mathrm{crit,full-c}}. \tag{24}$$

Equation (24) provides a lower limit on how much the critical time step can decrease in the worst case for a given $\alpha$, i.e., in the presence of the most unfavorable cut configuration. Based on this estimate, we define the modified CFL condition as

$$\Delta t \leq \alpha^{\frac{1}{3}}\, \Delta t_{\mathrm{crit,full-c}}. \tag{25}$$

We can further expand (25) by splitting the critical time step size of an uncut element into

$$\Delta t_{\mathrm{crit,full-c}} = C_{\mathrm{CFL}}(p)\, \frac{h}{c}, \tag{26}$$

where $0 < C_{\mathrm{CFL}} \leq 1$ is a constant depending on the problem's dimension, the time integration scheme, and the spatial discretization method (particularly its polynomial degree) [9]. Using this definition, we obtain the following equivalent CFL condition:

$$\Delta t \leq \alpha^{\frac{1}{3}}\, C_{\mathrm{CFL}}(p)\, \frac{h}{c}. \tag{27}$$

### Refinement of the estimate

Figure 6b shows that scaling the critical time step by $\alpha^{1/3}$ is often conservative. The true limit may be considerably larger, i.e., the ratio between $\Delta t_{\mathrm{crit,min}}$ and $\Delta t_{\mathrm{crit,full-c}}$ exceeds the idealized $\alpha^{\frac{1}{3}}$. This deviation initially increases when reducing $\alpha$, but soon converges to a constant offset. In the following, we augment (24) by a correction factor quantifying the difference to the true critical time step, allowing us to provide a more accurate estimate. We call this correction factor the scaling offset $k$, defined as

$$k(p, \alpha) = \frac{\Delta t_{\mathrm{crit,min}}}{\alpha^{\frac{1}{3}}\Delta t_{\mathrm{crit,full-c}}}. \tag{28}$$

Using (28), we obtain the modified lower-bound estimate

$$\Delta t_{\mathrm{crit,cfl,fcm-corr}} = k(p,\alpha)\,\Delta t_{\mathrm{crit,cfl,fcm}}. \tag{29}$$

Our refined CFL condition using the corrected critical time step then becomes

$$\Delta t \leq k(p,\alpha)\,\alpha^{\frac{1}{3}}\,\Delta t_{\mathrm{crit,full-c}} \tag{30}$$

or equivalently

$$\Delta t \leq k(p,\alpha)\,\alpha^{\frac{1}{3}}\,C_{\mathrm{CFL}}(p)\,\frac{h}{c}. \tag{31}$$

Figure 7 shows the values of $k$ for a range of polynomial degrees and dimensions $d = 1$ to $d = 3$. Moreover, Appendix C provides tabulated values of $k$ for a wide range of configurations. While the estimates are based on the setup of the previous section, our experience and the results shown later indicate that they hold in practice.

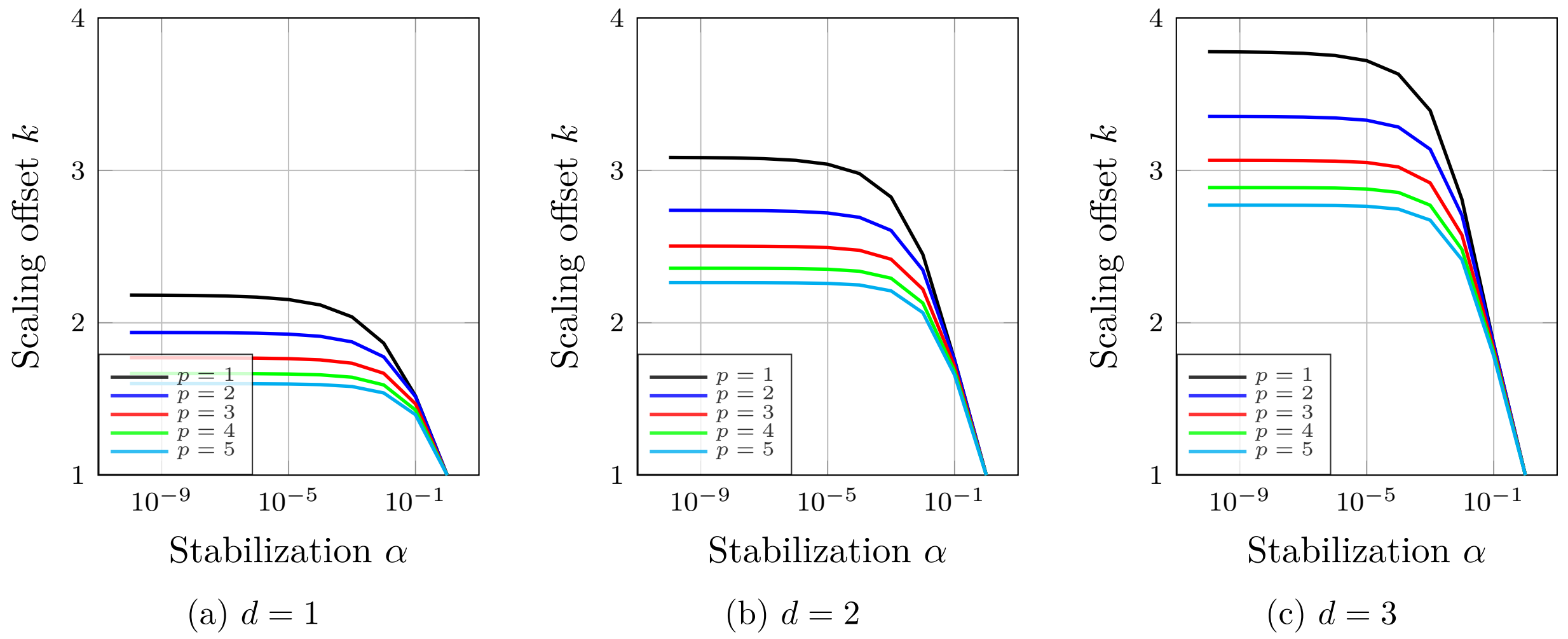

Figure 7: Scaling offsets for sliver cuts extracted from Figure 6b as the ratio between the results for different values of $p$ and the dashed reference line.

## Perforated plate study

In this section, we validate the proposed FCM-CFL condition on a perforated plate with the extended domain defined as $\Omega^e = [0,4] \times [0,2]$. The geometry features three vertical columns of circular holes with uniform radii $r_i = r = 0.2$, and equal horizontal and vertical spacing of $3 \cdot r = 0.6$. The central column is symmetrically aligned with the horizontal centerline of the plate. We assume unit density and wave speed $\rho = c = 1$, and impose homogeneous (reflecting) Neumann boundary conditions on $\partial\Omega$. As the critical time step size is governed solely by the eigenvalues of the mass and stiffness matrices, no external excitation is applied. The SCM-discretization employs a regular grid of $60 \times 30$ elements, stabilized with $\alpha = 10^{-4}$, yielding an element size of $h = \frac{1}{15}$. Fictitious (empty) elements are excluded from the computational domain. Uncut elements are integrated using $p+1$ GLL points per spatial direction to obtain a diagonal mass matrix, and $p+1$ Gauss-Legendre (GL) points per spatial direction for the stiffness matrix. For cut elements, the integration domain is refined via a quadtree with $k = p+3$ refinement levels towards the physical boundary. Each leaf of the quadtree employs $p+1$ GL points to integrate both the element mass and stiffness matrices.

Using this setup, we investigate the critical time step size by slightly shifting the circle positions in the $x$- and $y$-directions up to maximum distances of $\Delta x_{\max} = h$ and $\Delta y_{\max} = 2\,r$. To generate a wide variety of cut geometries, we sample $\Delta x_{\max}$ using 15 steps and $\Delta y_{\max}$ using 100 steps, resulting in 1 500 distinct configurations. As the holes are shifted vertically, the top row of circles crosses the upper boundary, while new circles enter from below. Figure 8 shows configurations #1 and #701 to illustrate the setup. Figure 9 shows the critical time step sizes for all 1 500 configurations for polynomial degrees $p = 1$ to $p = 5$. One dot corresponds to the simulation of one cut configuration. The black dots show the smallest critical time step size among the elements of that configuration ($\Delta t_{\text{crit,element}}$), while the purple dots represent the critical time step sizes of the resulting globally assembled system ($\Delta t_{\text{crit,global}}$). The solid red lines show the estimated lower bound for the critical time step size ($\Delta t_{\text{crit,cfl,fcm}}$) based on (24):

$$\Delta t_{\text{crit,cfl,fcm}} = \left(10^{-4}\right)^{\frac{1}{3}} \Delta t_{\text{crit,full-c}}. \tag{32}$$

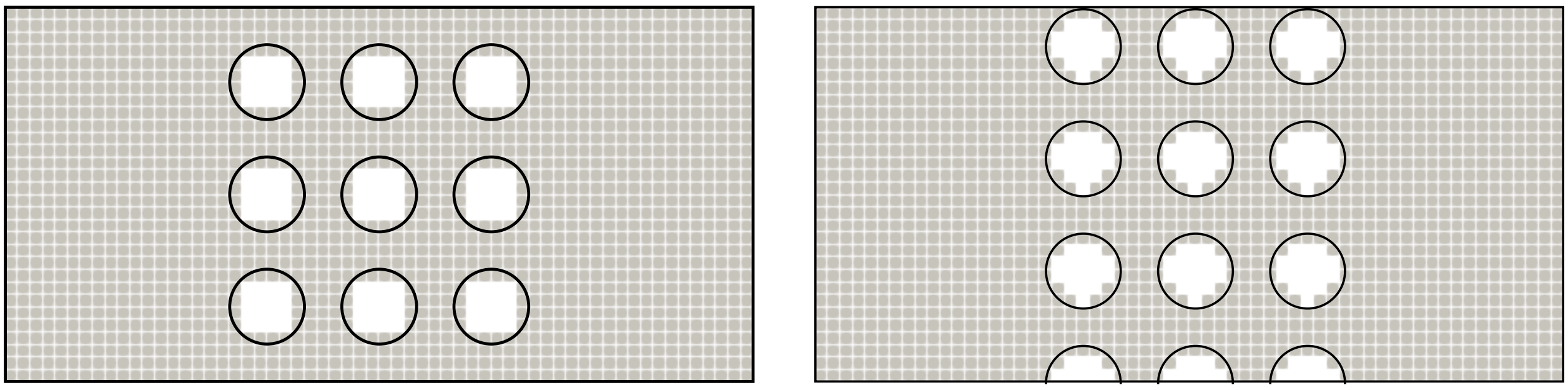

Figure 8: Perforated plate with computational mesh: Configuration #1 with $\Delta x = 0.0$ and $\Delta y = 0.0$ (left); Configuration #701 with $\Delta x = 0.033333$ and $\Delta y = 0.18990$ (right).

The dashed red lines show the refined estimate ($\Delta t_{\text{crit,cfl,fcm-corr}}$) based on (28):

$$\Delta t_{\text{crit,cfl,fcm-corr}} = k(\alpha = 10^{-4}, p)\, \Delta t_{\text{crit,cfl,fcm}}, \tag{33}$$

where $k(\alpha = 10^{-4}, p)$ are taken from Table 3. The green lines indicate the critical time step sizes of an element computed with a consistent (exactly integrated) mass matrix ($\Delta t_{\text{crit,full-c}}$). Finally, the blue lines indicate the critical time step sizes of an uncut element using nodal lumping for the mass matrix ($\Delta t_{\text{crit,full-l}}$).

The example confirms, that (25) provides a conservative lower bound of the critical time step size, while (30) refines the estimate by using the correction factor $k(\alpha, p)$. Across all configurations and polynomial degrees, the element-wise minimum critical time step sizes consistently lie just above the estimated limit. Effectively, $\alpha$-stabilization limits the smallest element-wise critical time step sizes to approximately one tenth of $\Delta t_{\text{crit,full-c}}$ for the investigated example. Since the largest eigenvalue of the global system is always smaller than or equal to the largest local eigenvalue among all elements [46, 47], the proposed estimate also bounds the global critical time step size. The two purple outliers for $p = 2$ and $p = 4$ correspond to configuration #701, for which the eigensolver did not converge. However, the corresponding simulations were stable, confirming that the stability limit still holds.

While the stabilized critical time step is still restrictive, we emphasize that even in boundary-conforming discretizations, the critical time step size can be significantly below the value expected based on the target element size. In particular, conforming the mesh to small geometric features, as would be necessary in configuration #701 (Figure 8), results in highly distorted, small elements, which also lead to severely reduced critical time step sizes.

## 6. Conclusion

This paper investigates the challenge of restrictive critical time step sizes in explicit time integration arising from badly cut elements in immersed boundary methods. Applying the finite cell method, the addition of non-zero $\alpha$-values in the fictitious domain prevents the mass and stiffness of the discretized system from reaching zero, thereby imposing a lower limit on the critical time step size. We first analytically study a linearly-interpolated single-element example considering two types of cut configurations: corner and sliver cuts. The analysis shows that the critical time step size does not decrease unboundedly with the cut ratio, but reaches a minimum that depends on the stabilization parameter $\alpha$. Specifically, we show that this minimum scales with $\alpha^{\frac{1}{3d}}$ for corner cuts and $\alpha^{\frac{1}{3}}$ for sliver cuts. Sliver cuts therefore pose a more severe limitation, as their critical time step size is independent of the spatial dimension. For sliver cuts, we then extend our analysis to high-order polynomials, where we numerically study the dependence of the critical time step size on the polynomial degree. Here, $\alpha$-stabilization again imposes a lower limit on the critical time step size. We show that the adverse effects of cut elements only weakly depend on the polynomial degree. Based on this insight, we propose a modified CFL condition tailored to FCM, which provides a conservative estimate of the critical time step size by scaling the critical time step of the underlying base discretization by $\alpha^{\frac{1}{3}}$. The CFL condition can be refined using a correction factor $k(\alpha, p)$, leading to a sharper estimate of the critical time step size. Tabulated values of $k(\alpha, p)$ are provided in the appendix. We demonstrate the validity of our CFL condition by considering wave propagation in a two-dimensional perforated plate comprising 1 500 model configurations with varying cut configurations. When combined with HRZ lumping or a weak enforcement of Dirichlet boundary conditions, $\alpha$-stabilization does not limit the critical time step size. Future work will therefore focus on finding ways to stabilize weakly imposed Dirichlet boundary conditions.

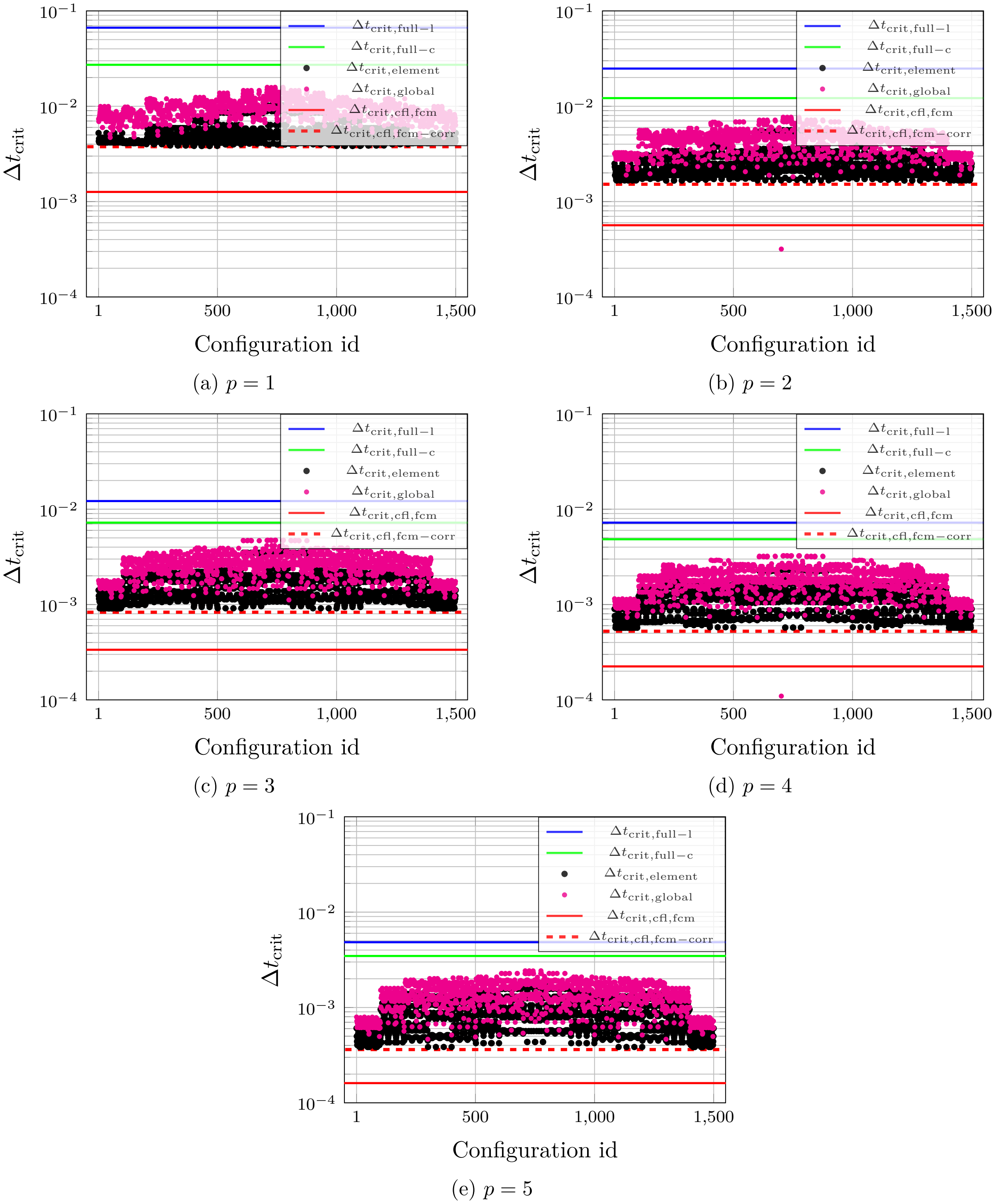

(a) $p = 1$

(b) $p = 2$

(c) $p = 3$

(d) $p = 4$

(e) $p = 5$

Figure 9: Critical time step sizes of the perforated plate with $60 \times 30$ elements and $\alpha = 10^{-4}$.

For choosing $\alpha$ in practice, one must not only relate it to $\Delta t_{\text{crit}}$, but also to a measure of the consistency error. In this context, we mention the analysis provided by [18], which found that the finite cell method introduces an error in the energy norm proportional to $\sqrt{\alpha}$ for Laplace and Helmholtz problems. We expect that for explicit dynamics, a relatively high stabilization, such as $\alpha = 10^{-4}$, retains sufficient accuracy for many engineering applications while permitting moderate minimum critical time step sizes.

## A. Asymptotic minimum critical time step size of the single-DOF system for a corner cut

To find the largest eigenvalue for our single DOF system for a corner cut we aim to maximize

$$\lambda = 3\,d\,\frac{(1-\alpha)\,\chi^{3d-2} + \alpha}{(1-\alpha)\,\chi^{3d} + \alpha} \tag{34}$$

for a given $\alpha$ (see Table 1). To this end, we require the partial derivative with respect to the cut position to vanish:

$$\frac{\partial \lambda}{\partial \chi} = \frac{3\chi^{3d-3} d(\alpha-1)(3\alpha\chi^2 d - 2\alpha\chi^{3d} - 3\alpha d + 2\alpha + 2\chi^{3d})}{((1-\alpha)\,\chi^{3d} + \alpha)^2} = 0.$$

For the relevant parameter range, this expression simplifies to

$$3\alpha\chi^2 d - 2\alpha\chi^{3d} - 3\alpha d + 2\alpha + 2\chi^{3d} = 0$$

or

$$\alpha(3\chi^2 d - 2\chi^{3d} - 3d + 2) + 2\chi^{3d} = 0.$$

It is clear that when asymptotically reducing $\alpha \to 0$, also $\chi$ must approach zero. Since in this case, the term multiplying $\alpha$ converges to $-3d + 2$, we know that the cut position maximizing (34) asymptotically approaches

$$2\chi^{3d} = \alpha(3d-2) \qquad \Longrightarrow \chi = \left(\alpha\frac{3d-2}{2}\right)^{\frac{1}{3d}}.$$

Substituting this result into (34) yields the following asymptotic estimate for $\alpha \to 0$

$$\begin{aligned}
\lambda_{\max} &\sim 3d\,\frac{(1-\alpha)\,\left(\alpha\frac{3d-2}{2}\right)^{\frac{3d-2}{3d}} + \alpha}{(1-\alpha)\,\left(\alpha\frac{3d-2}{2}\right)^{\frac{3d}{3d}} + \alpha} = 3d\,\frac{(1-\alpha)\,\left(\frac{3d-2}{2}\right)^{1-\frac{2}{3d}}\alpha^{1-\frac{2}{3d}} + \alpha}{\frac{3d}{2}\alpha - \frac{3d-2}{2}\alpha^2} \\
&\sim 3d\,\frac{\left(\frac{3d-2}{2}\right)^{1-\frac{2}{3d}}\alpha^{1-\frac{2}{3d}}}{\frac{3d}{2}\alpha} \\
&\sim C_\lambda(d)\,\alpha^{-\frac{2}{3d}},
\end{aligned}$$

with

$$C_\lambda(d) = 2\left(\frac{3d-2}{2}\right)^{1-\frac{2}{3d}}.$$

Using (11), we obtain the minimum time step size for $\alpha \to 0$:

$$\Delta t_{\text{crit, min}} \sim \frac{2}{\sqrt{C_\lambda(d)}}\,\alpha^{\frac{1}{3d}}.$$

# B. Asymptotic minimum critical time step size of the single-DOF system for a sliver cut

For sliver cuts, we instead search for the maximum of

$$\lambda = 3\,\frac{(1-\alpha)\left((d-1)\,\chi^3+\chi\right)+d\,\alpha}{(1-\alpha)\chi^3+\alpha}, \tag{35}$$

for a given $\alpha$ (see Table 1). Searching for a stationary partial derivative yields

$$\frac{\partial\lambda}{\partial\chi} = 3\frac{(1-\alpha)((1-\alpha)(\alpha+\chi^3(1-\alpha))(3\chi^2(d-1)+1)-3\chi^2(1-\alpha)(\alpha^d+(1-\alpha)(\chi^3(d-1)+\chi)))}{((1-\alpha)\chi^3+\alpha)^2} = 0,$$

which simplifies to

$$\alpha(3(1+\alpha)\chi^5+(3\alpha-1)\chi^3+3(2\alpha d-\alpha+d)\chi^2+1)-2\chi^3.$$

We see that $\alpha \to 0$ again implies $\chi \to 0$, for which the term multiplying $\alpha$ approaches 1. Thus, we obtain the following asymptotic limit for the cut position maximizing (35):

$$2\chi^3 = \alpha \qquad \implies \chi = \left(\frac{\alpha}{2}\right)^{\frac{1}{3}}.$$

Substituting the cut position into (35) yields the following asymptotic estimate for $\alpha \to 0$

$$\begin{aligned}
\lambda_{\max} &\sim 3\frac{(1-\alpha)((d-1)\frac{\alpha}{2}+\left(\frac{\alpha}{2}\right)^{\frac{1}{3}})+\alpha d}{(1-\alpha)\frac{\alpha}{2}+\alpha} = 3\frac{(1-\alpha)(\frac{d-1}{2}+2^{-\frac{1}{3}}\alpha^{-\frac{2}{3}})+d}{\frac{3}{2}-\frac{\alpha}{2}} \\
&\sim 3\frac{2^{-\frac{1}{3}}\alpha^{-\frac{2}{3}}}{\frac{3}{2}} \\
&\sim C_\lambda(d)\,\alpha^{-\frac{2}{3}},
\end{aligned}$$

with

$$C_\lambda(d) = 2^{\frac{2}{3}}.$$

Finally, the minimum time step size for $\alpha \to 0$ is

$$\Delta t_{\text{crit, min}} \sim \frac{2}{\sqrt{C_\lambda(d)}}\,\alpha^{\frac{1}{3}} = 2^{\frac{2}{3}}\alpha^{\frac{1}{3}}.$$

## C. Tables with FCM-CFL scaling offset factor $k(p, \alpha)$ for $d = 1, 2, 3$

Table 2: Scaling constants $k$ for $d = 1$

| $\alpha \setminus p$ | 1 | 2 | 3 | 4 | 5 |
|---|---|---|---|---|---|
| $10^{0}$ | 1.0000 | 1.0000 | 1.0000 | 1.0000 | 1.0000 |
| $10^{-1}$ | 1.5203 | 1.5156 | 1.4678 | 1.4274 | 1.3965 |
| $10^{-2}$ | 1.8661 | 1.7761 | 1.6683 | 1.5920 | 1.5394 |
| $10^{-3}$ | 2.0384 | 1.8746 | 1.7345 | 1.6427 | 1.5816 |
| $10^{-4}$ | 2.1169 | 1.9111 | 1.7565 | 1.6583 | 1.5940 |
| $10^{-5}$ | 2.1523 | 1.9254 | 1.7645 | 1.6637 | 1.5980 |
| $10^{-6}$ | 2.1684 | 1.9315 | 1.7678 | 1.6657 | 1.5994 |
| $10^{-7}$ | 2.1759 | 1.9343 | 1.7691 | 1.6665 | 1.6000 |
| $10^{-8}$ | 2.1794 | 1.9355 | 1.7698 | 1.6669 | 1.6003 |
| $10^{-9}$ | 2.1809 | 1.9360 | 1.7701 | 1.6671 | 1.6003 |
| $10^{-10}$ | 2.1817 | 1.9364 | 1.7701 | 1.6671 | 1.6004 |

Table 3: Scaling constants $k$ for $d = 2$

| $\alpha \setminus p$ | 1 | 2 | 3 | 4 | 5 |
|---|---|---|---|---|---|
| $10^{0}$ | 1.0000 | 1.0000 | 1.0000 | 1.0000 | 1.0000 |
| $10^{-1}$ | 1.7567 | 1.7530 | 1.7155 | 1.6828 | 1.6573 |
| $10^{-2}$ | 2.4485 | 2.3459 | 2.2202 | 2.1296 | 2.0664 |
| $10^{-3}$ | 2.8247 | 2.6057 | 2.4169 | 2.2923 | 2.2093 |
| $10^{-4}$ | 2.9794 | 2.6921 | 2.4758 | 2.3383 | 2.2481 |
| $10^{-5}$ | 3.0406 | 2.7206 | 2.4936 | 2.3513 | 2.2586 |
| $10^{-6}$ | 3.0659 | 2.7311 | 2.4996 | 2.3553 | 2.2616 |
| $10^{-7}$ | 3.0770 | 2.7354 | 2.5018 | 2.3567 | 2.2626 |
| $10^{-8}$ | 3.0820 | 2.7371 | 2.5028 | 2.3574 | 2.2631 |
| $10^{-9}$ | 3.0842 | 2.7379 | 2.5032 | 2.3576 | 2.2632 |
| $10^{-10}$ | 3.0853 | 2.7384 | 2.5033 | 2.3577 | 2.2633 |

Table 4: Scaling constants $k$ for $d = 3$

| $\alpha \setminus p$ | 1 | 2 | 3 | 4 | 5 |
|---|---|---|---|---|---|
| $10^{0}$ | 1.0000 | 1.0000 | 1.0000 | 1.0000 | 1.0000 |
| $10^{-1}$ | 1.8639 | 1.8610 | 1.8308 | 1.8041 | 1.7830 |
| $10^{-2}$ | 2.8097 | 2.7055 | 2.5759 | 2.4810 | 2.4140 |
| $10^{-3}$ | 3.3925 | 3.1385 | 2.9177 | 2.7714 | 2.6734 |
| $10^{-4}$ | 3.6317 | 3.2843 | 3.0223 | 2.8554 | 2.7459 |
| $10^{-5}$ | 3.7199 | 3.3292 | 3.0519 | 2.8779 | 2.7645 |
| $10^{-6}$ | 3.7540 | 3.3442 | 3.0609 | 2.8843 | 2.7695 |
| $10^{-7}$ | 3.7684 | 3.3501 | 3.0640 | 2.8863 | 2.7711 |
| $10^{-8}$ | 3.7747 | 3.3523 | 3.0653 | 2.8872 | 2.7717 |
| $10^{-9}$ | 3.7774 | 3.3533 | 3.0658 | 2.8875 | 2.7718 |
| $10^{-10}$ | 3.7787 | 3.3539 | 3.0659 | 2.8875 | 2.7719 |

## Acknowledgements

We gratefully acknowledge the Deutsche Forschungsgemeinschaft (DFG, German Research Foundation) for their support through the grants KO 4570/1-2 and RA 624/29-2 (both grant number 438252876) as well as DU 405/20-1 (grant number 503865803). We would also like to thank Alexander Düster, Stefan Kollmannsberger, and Ernst Rank for many fruitful discussions on this topic that inspired this study. Lastly, we would like to thank the Computational Mechanics reviewers for their remarkably detailed feedback on this contribution.